\documentclass[a4paper,12pt]{article}

\title{A necessary condition for brane solutions to have their origins in string field theory}
\author{Riuji Mochizuki\thanks{e-mail address:rjmochi@tdc.ac.jp} and Kenji Ikegami\thanks{e-mail address:kikegami@tdc.ac.jp}
 \\  {\small  Laboratory of Physics, Tokyo Dental College, Chiba 261-8502, Japan}}

\date{\today}

\begin{document}

\maketitle
\abstract{
We present a necessary condition for brane solutions to have their origins in string field theory. For such solutions, T-duality
 operations between IIA and IIB solutions should be well-defined.  Nevertheless, not all S-brane solutions have T-duality.  
The solutions that have T-duality and may accordingly be regarded as low-energy 
 solutions of string field theory satisfy a condition in which some S-brane distributions are forbidden.


 }
\section{Introduction}

In many brane-world models\cite{ngreview}, our 4-dimensional spacetime is thought to be 
described as a low-energy approximation of a solution of string field theory.  Unfortunately, however, 
since we do not understand string field theory well, we cannot conclude that a given supergravity solution has its origin in 
string field theory.   
If it is a low-energy solution of string theory, dimensional reduction (or oxidation) and T-duality
 operations should be well-defined.   Dimensional reduction and T-duality 
 operation rules are given for time-independent brane solutions \cite{Ortin}.  Nevertheless, the situation is rather different when we consider time-dependent solutions.  Although
the dimensional reduction of spacelike brane solutions has been discussed \cite{Roy}\cite{ohtan}, it is not well-defined at all times.
  Furthermore, their T-duality is hardly discussed in any papers.  It is the time-dependence of the dilaton expectation values
   that makes this problem so complicated. 
 If an S-brane solution does not have T-duality, we should not regard it as a low-energy solution of string theory.  
 
In our previous papers\cite{mochi}\cite{mochi2}\cite{mi1}, we constructed singular (i.e. neither general nor particular) S-brane solutions of supergravity.
  Furthermore, most of them possess static, flat dimensions.  
This is of merit, as it enables us to define the dimensional reduction and T-duality of time-dependent solutions.  
In this paper, we examine these solutions and find that compactification of dimensions whose metric tensors depend on time is ill-defined and 
that the time-dependence of 
dilaton expectation values spoils T-duality.  
These observations lead us to conclude that singular solutions of 10-dimensional supergravity solutions should be regarded as low-energy solutions of
string field theory
 only if the dilaton expectation values are independent of time.  More generally, we present a necessary condition for general brane solutions 
 which should be satisfied if they are to be regarded as low-energy solutions of string field theory.

The organization of this paper is as follows: 
our starting point is Einstein gravity coupled to a dilaton and $n$-form fields in 11 dimensions (M-theory) and 10 dimensions (IIA and 
IIB superstring theories).  In section 2 we write a singular, spacelike brane solution by following \cite{mochi2}\cite{mi1}.  
In section 3, the T-duality of these solutions is discussed and our conclusion derived.

\section{Singular spacelike solutions}

Firstly, we present singular S-brane solutions according to our previous papers\cite{mochi}\cite{mochi2}\cite{mi1}.  
We consider Einstein gravity coupled to a dilaton field $\phi$ and  $m$ kinds of {\it n}-form field $F_{n}$, whose action $I$ is

\begin{equation}
I={1\over 16\pi G}\int d^Dx\surd \overline{- g}\Big[ R-{1\over 2}g^{\mu\nu}\partial_{\mu} \phi\partial_{\nu}\phi
-\sum_{A=1}^{m}{1\over 2\cdot n_A!}e^{\alpha_A\phi}F_{n_A}^{\ 2}\Big],\label{eq:action}
\end{equation}
where $\alpha_A$ is the dilaton coupling constant given by
\[
\alpha_A = \left\{
\begin{array}{cl}
0 & ({\rm M-theory}) \\
-1 & ({\rm NS-NS\  sector}) \\
{5-n_A\over 2} & ({\rm R-R\  sector})
\end{array}
\right.
\]
and {\it D}=11
 for M-theory and {\it D}=10 for superstring theories.  

We can write a set of solutions of the field equations and the Bianchi identity.  These solutions may include 
S{\it w} and S{\it m}, which are spacelike counterparts of wave and monopole solutions, respectively. 
  The metric form is 
\begin{equation}
ds^2 = H_1du^2 + H_2dv^2 + \sum_{i=3}^{p+1}H_idx^idx^i+\sum_{a,b=p+2}^D H_0\eta_{ab}dy^ady^b,\label{eq:setting}
\end{equation}
where
\begin{equation}
\eta_{ab} = \{diag.(+ , \cdots , + , - )\},\label{eq:singletime}
\end{equation}
\[
du = dx^1 + 2i\delta_m\tilde B_ady^a,
\]
\[
dv = dx^2 + i\delta_w(H-1)dx^3,
\]
with
\[
\delta_{m(w)} = \left\{
\begin{array}{cl}
1 & ({\rm S}m(w){\rm \  is\  included.}) \\
0 & ({\rm S}m(w){\rm \  is\ not\  included.})
\end{array},
\right.
\]
\[
\partial_a\tilde{B}_{b}-\partial_b\tilde{B}_{a}=\eta_{ac}\eta_{bd}\epsilon^{cde}H^{-2}\partial_eH,
\]
\begin{equation}
\partial^2H^{-1}(y)\equiv\eta^{ab}\partial_a\partial_bH^{-1}(y)=0.\label{eq:hh}
\end{equation}
We use $\{ x^i;\ i=1,\cdots,p+1\}$ as the coordinates of the space where the branes exist.   
General orthogonally intersecting S{\it p}-brane solutions have been given \cite{gen}, where the metric functions 
 and fields depend only on $y^D$.   A D-brane solution which depends on all the extra space coordinates 
 has been suggested\cite{multi}.
 
The metric functions are
\begin{equation}
H_1(y)=H^{\delta_m + \sum_{A=1}^m{\delta_{A,i}\over D-2}}(y),\label{eq:solu1}
\end{equation}
\begin{equation}
H_2(y)=H^{-\delta_w + \sum_{A=1}^m{\delta_{A,i}\over D-2}}(y),\label{eq:solu2}
\end{equation}
\begin{equation}
H_3(y)=H^{\delta_w + \sum_{A=1}^m{\delta_{A,i}\over D-2}}(y),\label{eq:solu3}
\end{equation}
\begin{equation}
H_i(y)=H^{\sum_{A=1}^m{\delta_{A,i}\over D-2}}(y)\ \ \ \ (i=4, \cdots ,p+1),\label{eq:solu4}
\end{equation}
\begin{equation}
H_0(y)=H^{-\delta_m -\sum_{A=1}^m{q_A+1\over D-2}}(y),\label{eq:solv}
\end{equation}
where
\begin{equation}
\delta_{A,i} = \left\{
\begin{array}{cl}
D-q_A-3 & (i\in q_A) \\
-(q_A+1) & (i\notin q_A)
\end{array}
\right..\label{eq:deltai}
\end{equation}

The field strength for an electrically charged S{\it q}-brane is given by
\begin{equation}
(F_n)_{i_{1}\cdots i_{n-1}a}(y)=\epsilon_{i_1\cdots i_{n-1}}\partial_a E(y),
\end{equation}
\[
(n = q+2),
\]
while the magnetically charged case is given by
\begin{equation}
(F_n)^{a_1\cdots a_{n}}(y)=\frac{1}{\surd \overline{-g}}e^{-\alpha\phi}\epsilon^{a_1\cdots a_{n}b}\partial_b E(y),
\end{equation}
\[
(n = D-q-2),
\]
where
\begin{equation}
E(y)=iH(y).\label{eq:solE}
\end{equation}
The dilaton field is
\begin{equation}
\phi(y) =H^{-\sum_{A=1}^m{\varepsilon_A\alpha_A\over 2}}(y),\label{eq:solphi}
\end{equation}
where
\[
\varepsilon_A = \left\{
\begin{array}{cl}
+1 & (F_{n_A}\ {\rm is\ an\ electric\ field\ strength}) \\
-1 & (F_{n_A}\ {\rm is\ a\ magnetic\ field\ strength})
\end{array}
\right..
\]
For S{\it w} and S{\it m}, the dilaton coupling constant $\alpha_A=0$.

The intersection rule, which 
has been suggested for general solutions in other papers \cite{gen}\cite{int1}\cite{int2}, should be satisfied:
\begin{equation}
-\varepsilon_A\varepsilon_B\alpha_A\alpha_B-2(\bar q+1)+{2(q_A+1)(q_B+1)\over D-2}=0,\label{eq:intersection}
\end{equation}
where $\bar q+1$ is the number of dimensions which $q_A$-brane and $q_B$-brane are crossing on.  
An S{\it w} can be put in any two isometric directions, while an S{\it m} needs a direction with no other branes.

To discuss dimensional reduction and T-duality, we construct a singular solution with static dimensions.
To this end, we consider solutions which depend only on the scale parameter $r$ of the entire, or a part of,  spacetime perpendicular to the brane:
\begin{equation}
r \equiv \sqrt{-\eta_{ab}y^{a}y^{b}},\ \ \ \ -\eta_{ab}y^{a}y^{b}>0,
\end{equation}
Note that $r$ is a timelike coordinate.  To satisfy (\ref{eq:hh}),
\begin{equation}
H = r^{D-p-3}.\label{eq:HH}
\end{equation}
Then, the metric of this spacetime becomes
\begin{equation}
ds^2=\sum_{i=1}^{p+1}H_idx^idx^i-H_0\Big(dr^2 - r^2d\Sigma^2_{D-p-2} \Big), \label{eq:metric}
\end{equation}
after some coordinate transformation if S{\it w} and/or S{\it m} are included.  
 $d\Sigma_{D-p-2}$ is the line element of a $(D-p-2)$-dimensional hyperbolic space whose scale factor is unity.

We define cosmic time ({\it our time}) $t$ as
\begin{eqnarray}
dt&\equiv&H_0^{\ 1/2}dr\nonumber\\
&=&r^{(D-p-3)\big[-{\delta_m\over 2}-\sum{q_A+1\over 2(D-2)}\big]}dr,
\end{eqnarray}
and impose a condition:
\begin{equation}
(D-p-3)\big[{\delta_m\over 2}+\sum_{A=1}^m{q_A+1\over 2(D-2)}\big]=1.\label{eq:joukenn}
\end{equation}

In this case, since
\begin{equation}
t= \ln r,
\end{equation}
\begin{equation}
H_0=r^{-2}.
\end{equation}
\begin{equation}
H = {\rm e}^{(D-p-3)t},\label{eq:ht}
\end{equation}
the metric (\ref{eq:metric}) becomes
\begin{equation}
ds^2=-dt^2+\sum_{i=1}^{p+1}H_i(t)dx^idx^i+d\Sigma^2_{D-p-2}\ , \label{eq:metric2}
\end{equation}
where $H_i(t)$s are defined by substituting (\ref{eq:ht}) for $H(y)$ in (\ref{eq:solu1})$\sim$(\ref{eq:solv}).
Note that the scale factor of the extra space is independent of $t$ and 
some $H_i$s may be independent of time in this metric.

All the solutions satisfying the above condition (\ref{eq:joukenn}) and possessing static flat 
dimensions are given in \cite{mi1}. 

\section{A necessary condition for brane solutions }

In this section we discuss the T-duality of S-brane solutions and derive a necessary condition for them to have their origins in string field theory.  
We expect that the time-independent rule of dimensional reduction can be applied to static flat dimensions of S-brane solutions.  
In fact, we can easily confirm this.  Dimensional reduction is well-defined 
and other dimensions behave in exactly the same way as they did before dimensional reduction for all the static flat dimensions.  
For example, let us consider a $D=11$, $p=6$ solution in which 3 SM2 branes are located on the 1st, 2nd and 4th, 
on the 1st, 3rd and 5th and on the 2nd, 3rd and 6th dimensions.
Its metric is 
\begin{equation}
ds^2 = -dt^2 + \sum_{i=1}^3 {\rm e}^{2t}dx^idx^i + \sum_{j=4}^6dx^jdx^j + {\rm e}^{-2t}dx^7dx^7 + d\Sigma^2_3.\label{eq:example}
\end{equation}
We can compactify the 4th, 5th and 6th dimensions.  When the number of the spacetime dimensions is
 reduced from 11 to 8, the metric of the 8 remaining dimensions stays as it was in 11 dimensional spacetime:
\begin{equation}
ds^2 = -dt^2 + \sum_{i=1}^3 {\rm e}^{2t}dx^idx^i +  {\rm e}^{-2t}dx^7dx^7 + d\Sigma^2_3.\label{eq:remain}
\end{equation}
If only the 4th dimension is compactified, we obtain
\begin{equation}
ds^2 = -dt^2 + \sum_{i=1}^3 {\rm e}^{2t}dx^idx^i + \sum_{j=5}^6dx^jdx^j + {\rm e}^{-2t}dx^7dx^7 + d\Sigma^2_3.
\end{equation}
This is a IIA supergravity solution which consists of 1 SF1 brane and 2 SD2 branes.  Similarly, all the $D=11$ M-theory solutions 
which satisfiy (\ref{eq:joukenn}) and have static dimensions become $D=10$ IIA supergravity solutions\cite{mi1} by dimensional reduction.
 Oxidation is naturally defined as the inverse operation of dimensional reduction.   
  On the other hand, if we compactify a time-dependent dimension (one of the 1st, 2nd, 3rd and 7th dimensions), metric functions of 
  the other dimensions change in accordance with reduction of the dimensions.  
We, therefore, conclude that reduction of dimensions which depends
  on time cannot be defined.

Using dimensional reduction, we can define T-duality between IIA and IIB solutions.  If IIA and IIB solutions are reduced to the same 9-dimensional
 solution, the IIA solution 
is the T-dual of the IIB solution and vice versa.  Obeying these rules,  we confirm that each IIA solution that satisfies (\ref{eq:joukenn})
and has static dimensions can find a IIB solution as its T-dual partner.  Nevertheless, some of such IIB solutions have no T-dual partners.  An example is
a $p=5$ IIB solution in which 2 SD1 branes are located on the 1st and 3rd and on the 2nd and 4th dimensions, and 1 SD3 brane on
 the 1st, 2nd, 5th and 6th dimensions.  Its metric is
\begin{equation}
ds^2 = -dt^2 + \sum_{i=1}^2 {\rm e}^{2t}dx^idx^i + \sum_{j=3}^6dx^jdx^j + d\Sigma^2_3.\label{eq:dame}
\end{equation}
If the 3rd dimension is compactified, its metric becomes
\begin{equation}
ds^2 = -dt^2 + \sum_{i=1}^2 {\rm e}^{2t}dx^idx^i + \sum_{j=4}^6dx^jdx^j + d\Sigma^2_3.
\end{equation}
No IIA solutions are, however, obtained from this $D=9$ solution by oxidation. That is, (\ref{eq:dame}) has no
T-dual partner.

It is the dilaton expectation values that cause this difference.  
   The T-duality of a solution is well-defined 
if its dilaton expectation value is independent of time.  If not, T-duality does not exist for the solution.
 We are convinced of this reason if we examine T-duality operations in the Einstein frame.  A formula for 
converting a IIB solution to IIA by compactifying the $z$ direction is
\begin{eqnarray}
g^{A}_{\mu\nu}&=&(g^B_{zz})^{\frac{2}{D-2}} \exp\big\{\frac{8}{(D-2)^2}\phi^B\big\}\nonumber \\ 
&&\times\Big[
                g^B_{\mu\nu}
                -\frac{g^B_{z\mu}g^B_{z\nu}-4B^{(1)}_{z\mu}B^{(1)}_{z\nu}\exp\{-\frac{8}{D-2}\phi^B\}}
                  {g^B_{zz}}\Big],\label{eq:convert}
\end{eqnarray}
where $g^A$ and $g^B$ are IIA and IIB metric tensors, respectively.  If a IIB metric tensor in (\ref{eq:convert}) is static and $\phi^B$ depends on time, 
a IIA metric tensor must depend on time.  That is, there exists no IIA solution that can be reduced to the same $D=9$ solution that this IIB solution is reduced to.  
Thus, we argue that 
such  a IIB solution has no T-duality.  
A constant dilaton expectation value in 10-dimensional spacetime corresponds to a constant coupling in our 4-dimensional spacetime.  
This condition is agreeable and  sometimes imposed by hand on phenomenological models.

Moreover, taking account of (\ref{eq:solphi}) and assuming that string field theory would determine 
brane distribution, we conclude that supergravity solutions should satisfy a condition
\begin{equation}
\sum^{m}_{A=1}\varepsilon_A\alpha_A = 0,\label{eq:necessary}
\end{equation}
if they have their origins in string field theory.  We hypothesise that 
this necessary condition also applies to general S-brane solutions, 
even if no dimensions can be reduced.  


\
\

\

We would like to thank Associate Professor Jeremy Williams, Tokyo Dental College, for his assistance 
with the English of this manuscript.

\end{document}